\documentclass[twocolumn]{bmcart}

\usepackage[utf8]{inputenc} 

\usepackage{amsthm,amsmath,bm,mathtools,amssymb}
\usepackage{algorithmic}
\usepackage{algorithm}

\usepackage{amsmath}
\usepackage[cmintegrals]{newtxmath}
\usepackage{bm}
\usepackage{cite}
\usepackage{booktabs}
\usepackage{multirow}
\usepackage{stfloats}
\usepackage{array} 
\usepackage[caption=false]{subfig}

\def\includegraphics{}

\startlocaldefs
\endlocaldefs

\begin{document}

\begin{frontmatter}

\begin{fmbox}
\dochead{Research}


\title{Learning-based $A$ $Posteriori$ Speech Presence Probability Estimation and Applications}   


\author[
   addressref={aff1},                   
  corref={aff1},                       
   email={stao@es.aau.dk}   
]{\inits{ST}\fnm{Shuai} \snm{Tao}}
\author[
   addressref={aff1},
   email={jrj@es.aau.dk}
]{\inits{JRJ}\fnm{Jesper Rindom} \snm{Jensen}}
\author[
   addressref={aff2},
   email={xiangyang3131777@outlook.com}
]{\inits{YX}\fnm{Yang} \snm{Xiang}}
\author[
   addressref={aff1},
   email={himavanth.reddy19@gmail.com}
]{\inits{HR}\fnm{Himavanth} \snm{Reddy}}
\author[
   addressref={aff3},
   email={qzzhang@mail.nwpu.edu.cn}
]{\inits{QZZ}\fnm{Qingzheng} \snm{Zhang}}
\author[
   addressref={aff1},
   email={mgc@es.aau.dk}
]{\inits{MGC}\fnm{Mads Gr\ae sb\o ll } \snm{Christensen}}


\address[id=aff1]{%
  \orgname{Electronic Systems, Aalborg University},
  \street{Fredrik Bajers Vej 7}
  \postcode{9220}
  \city{Aalborg},
  \cny{DK}
}

\address[id=aff2]{
  \orgname{Hithink RoyalFlush AI Research Institute \& Tsinghua University}, 
  \street{No. 18 Tongshun Street, Wuchang Street},                     %
  \postcode{310023}                                
  \city{Hangzhou},                              
  \cny{China}                                    
}

\address[id=aff3]{
  \orgname{Northwestern Polytechnical University}, 
  \street{No. 127, Zhangjiacun Street},                     %
  \postcode{710072}                                
  \city{Xi'an},                              
  \cny{China}                                    
}





\begin{abstractbox}

\begin{abstract} 
The $a$ $posteriori$ speech presence probability (SPP) is the fundamental component of noise power spectral density (PSD) estimation, which can contribute to speech enhancement and speech recognition systems. Most existing SPP estimators can estimate SPP accurately from the background noise. Nevertheless, numerous challenges persist, including the difficulty of accurately estimating SPP from non-stationary noise with statistics-based methods and the high latency associated with deep learning-based approaches. This paper presents an improved SPP estimation approach based on deep learning to achieve higher SPP estimation accuracy, especially in non-stationary noise conditions. To promote the information extraction performance of the DNN, the global information of the observed signal and the local information of the decoupled frequency bins from the observed signal are connected as hybrid global-local information. The global information is extracted by one encoder. Then, one decoder and two fully connected layers are used to estimate SPP from the information of residual connection. To evaluate the performance of our proposed SPP estimator, the noise PSD estimation and speech enhancement tasks are performed. In contrast to existing minimum mean-square error (MMSE)-based noise PSD estimation approaches, the noise PSD is estimated by the sub-optimal MMSE based on the current frame SPP estimate without smoothing. Directed by the noise PSD estimate, a standard speech enhancement framework, the log spectral amplitude estimator, is employed to extract clean speech from the observed signal. From the experimental results, we can confirm that our proposed SPP estimator can achieve high noise PSD estimation accuracy and speech enhancement performance while requiring low model complexity.
\end{abstract}




\end{abstractbox}
\end{fmbox}

\end{frontmatter}



\section{Introduction}
The speech presence probability (SPP) estimator operates in the short time-frequency transform (STFT) domain, which provides the speech presence probability for each time-frequency (T-F) bin. Theoretically, the SPP is defined as the ratio of the likelihood functions of speech presence and speech absence \cite{ephraim1984speech, cohen2003noise, gerkmann2008improved}. Additionally, with background noise interference, the clean speech signal is commonly corrupted by additive noise, and most of the $a$ $posteriori$ SPP estimators are derived based on the assumption that the spectral coefficients of noise and speech can be modeled using complex Gaussian distributions and that they are independent \cite{middleton1968simultaneous, breithaupt2010analysis, gerkmann2010empirical}. Since SPP can softly determine whether speech is present under noise interference, it attracts great interest in integrating SPP with speech processing systems, such as mobile communication, voice analysis, and hearing aids \cite{ngo2009incorporating, nielsen2023analysis, zhang2024time}. 

In most existing SPP-based speech signal processing applications, SPP is employed to estimate noise power spectral density (PSD) \cite{chen2006new}. Since the noise PSD may change rapidly over time, it should be updated as frequently as possible in the noise estimation process. Conventional noise PSD estimation approaches work on the assumption that speech is not actually present, which relies on the detection of speech presence employing the hard voice activity detector (VAD) \cite{sohn1999statistical, ramirez2004voice, kim2018voice, zhu2023robust}. Nevertheless, under conditions of non-stationary noise and a low input signal-to-noise ratio (SNR), identifying noise through the hard VAD proves challenging. This difficulty arises because the system may misinterpret a sudden rise in noise PSD as the commencement of speech \cite{mckinley1997model, meyer1997comparison}. Compared with the hard VAD, a hard-decision speech presence detector, SPP provides a soft decision that can detect low-energy speech components, especially in non-stationary noise conditions \cite{cohen2001spectral, cohen2001speech}. As a result, when the SPP estimator reliably identifies speech presence, the distortion of the clean speech spectrum can be avoided in many aspects, such as noise reduction-based applications.

Most existing statistical model-based SPP estimators aim to achieve accurate SPP estimates for speech presence or speech absence with low latency and low computational complexity. In \cite{middleton1968simultaneous, ephraim1984speech, mcaulay1980speech}, the $a$ $posteriori$ SPP is computed by estimating the $a$ $priori$ SNR and the $a$ $posteriori$ SNR. However, these SPP estimators struggle to generate a low value for the T-F bin corresponding to speech absence, particularly during speech pauses and the harmonics of voiced speech. To overcome this problem, in \cite{cohen2001speech, cohen2002noise, cohen2003noise}, Cohen et al. proposed the $a$ $posteriori$ SPP estimator based on the $a$ $priori$ SPP controlled by the result of the minimum tracking. Since these SPP estimators are derived from the $a$ $priori$ SNR-parameterized likelihood functions, it is impossible to distinguish the presence or absence of speech when the $a$ $priori$ SNR is zero. In \cite{gerkmann2011unbiased, gerkmann2008improved}, Gerkmann et al. presented an optimally derived fixed value for the $a$ $priori$ SNR to implement a novel $a$ $posteriori$ SPP estimator, which achieved a better tradeoff between speech distortion and noise leakage. In addition, to reduce random fluctuations, the SPP estimator is derived from the smoothed observations. Nevertheless, these SPP estimators still have the drawback that they commonly assume the time and frequency of each T-F bin are independent when they estimate the SPP of each T-F bin. With this assumption, the performance of SPP estimation may be degraded since the spectral coefficients obtained after computing STFT are both correlated across time and frequency. Taking inter-band and inter-frame correlation into account, each T-F bin with its a few neighboring T-F bins is used to jointly estimate the SPP of each T-F bin to improve the speech detection accuracy further \cite{momeni2014single}. Additionally, using the knowledge of speech presence uncertainty in \cite{krawczyk2018speech}, Kim et al. proposed a speech power spectral density uncertainty based on the $a$ $posteriori$ SPP estimator \cite{kim2022improved}. 

Recently, with the development of deep learning, many novel SPP estimation methods based on deep neural networks (DNNs) have been proposed, and these estimators have been deployed successfully in speech signal processing applications, such as speech enhancement and speech recognition. To estimate SPP, the designed DNN model is commonly trained in a supervised manner using training data pairs. In \cite{tammen2020dnn} and \cite{wang2021multi}, the spectrum of the observed signal and the $a$ $posteriori$ SPP derived from \cite{gerkmann2011unbiased} were used as the input features and target, respectively. Since speech signal is a time-dependent series, recurrent neural networks are the critical component of the DNN model design. Since the mask ranges from 0 to 1, e.g., the ideal binary mask (IBM) and ideal ratio mask (IRM) meet the requirements of the SPP i.e., speech presence can be softly decided in each T-F bin by the IBM and IRM \cite{yang2018deep, tu2019speech, pfeifenberger2020resource, martin2020online}. In \cite{yang2018deep} and \cite{martin2020online}, the IBM-based SPP estimator was designed to estimate the IBM matrix as the SPP from the observed signal. When training the DNN-based IBM estimator, the target of the IBM is computed from the IRM with a pre-defined threshold i.e., the IRM value of the T-F bin set to 0 or 1 to represent speech absence or presence in the IBM matrix. Different thresholds exist in \cite{yang2018deep} and \cite{martin2018single} because the threshold is artificially set. Additionally, in \cite{martin2020online}, a DNN-based IBM estimator is only considered as the $a$ $priori$ SPP estimator, which is a key component of the $a$ $posteriori$ SPP estimator. To improve the robustness of the DNN-based SPP estimator in adverse environments, \cite{tu2019speech} proposed an improved SPP (ISPP) that incorporates the estimated IRM from a teacher model into the improved minima controlled recursive averaging (IMCRA) approach. In the multi-channel scenario, SPP is still an active research topic. To obtain the noise PSD matrix for MVDR beamforming and noise PSD for post-filtering, the $a$ $priori$ SNR estimate with a pre-defined threshold is regarded as the SPP in \cite{kim2023dnn}.

In terms of the DNN-based SPP estimator, there are some pros and cons. Although it can provide more accurate SPP estimation than the statistical model-based methods, it has higher latency and computational complexity. Additionally, since the data-driven method needs rich training data pairs to improve performance, most existing DNN-based SPP estimators rely on data volume and network depth. To overcome these problems, the IRM-based estimator in \cite{pfeifenberger2020resource} proposed a reduced-precision DNN that significantly reduces execution time and memory footprint. Moreover, in our previous work \cite{tao2023frequency}, we also focused on reducing the DNN-based SPP estimator model complexity, which can improve the efficiency of the estimator with high speech detection accuracy. Although in \cite{tao2023frequency}, decoupling frequency bins from the observed signal can reduce the model complexity sharply, we found that it may degrade the SPP estimation performance to some extent. Therefore, in \cite{tao2023single}, we proposed to concatenate the global information and local information to generate hybrid global-local information as the input features of the SPP decoder to estimate SPP, which achieved higher SPP estimation accuracy with low model complexity. However, there is still scope of improvement.

Hence, in this paper, we introduce one alternative high-performance decoder with residual connection to optimize the SPP estimation model in \cite{tao2023single}. Subsequently, to evaluate the performance of the SPP estimator, we deploy it to two downstream tasks, noise PSD estimation and speech enhancement, and provide a more in-depth analysis and evaluation of the learning-based $a$ $posteriori$ SPP estimator. More importantly, to help the learning-based SPP estimator better learn the data representation rather than a data stereotype, the actual value of the $a$ $priori$ SPP and SNR instead of the fixed value is used to estimate the $a$ $posteriori$ SPP during the training stage. Therefore, without any observations, we do not smooth the SPP estimate when updating it. With the SPP estimate, the sub-optimal MMSE is employed to estimate the noise PSD without smoothing. Finally, to reduce the noise estimation error influence, a robust SPP-based speech enhancement task is performed using the log spectral amplitude (LSA) estimator \cite{ephraim1985speech}. The paper is structured as follows. Section \ref{Signal Model} presents the signal model. Then, the related statistics-based $a$ $posteriori$ SPP estimation is discussed in Section \ref{statistics-based}. In Section \ref{learning-based model}, the learning-based SPP estimator is proposed, and the details of model architectures and training strategy are also presented. Implementation details and evaluation metrics are shown in Section \ref{implementation}. In Section \ref{resutls}, we deploy the proposed SPP estimator in noise PSD estimation and speech enhancement. Furthermore, we also discuss the observations in Section \ref{resutls}.

\section{Signal Model}\label{Signal Model}
 Firstly, at time instant $t$, considering that the clean speech $x(t)$ is corrupted by the ambient noise $n(t)$, recorded using a single microphone, the observed signal is given as
\begin{equation}
    y\left(t\right)=x\left(t\right)+n\left(t\right).
    \label{e1}
\end{equation}

Using the STFT analysis, in the T-F domain, the observed signal can be expressed by
\begin{equation}
     Y\left(k,l\right)=X\left(k,l\right)+N\left(k,l\right),
     \label{eq2}
\end{equation}
where $Y(k,l)$, $X(k,l)$, and $N(k,l)$ are the spectral coefficients for $y(t)$, $x(t)$, and $n(t)$, respectively. $k\in \{0,...,K-1 \}$ denotes frequency index, and $l\in \{0,...,L-1\}$ denotes frame index. For the rest of this paper, for brevity, we omit the frequency index $k$ as each frequency bin is considered separately unless otherwise indicated.

Assuming the spectral coefficients of the speech signal $X(l)$ and noise signal $N(l)$ follow zero-mean complex Gaussian distributions \cite{hendriks2010mmse} and there is no correlation between $X(l)$ and $N(l)$, the PSD of the observed signal $Y(l)$ can be obtained as
\begin{equation}
\begin{split}
    \Phi_{Y}(l)&=E\left[ |Y(l)|^2 \right]\\
    &=\Phi_{X}(l) + \Phi_{N}(l)
\end{split}
\label{eq3}
\end{equation}
where $E(.)$ is the statistical expectation operator, $\Phi_{X}(l)=E\left[|X(l)|^2 \right]$ is the speech PSD and $\Phi_{N}(l)=E\left[|N(l)|^2 \right]$ is the noise PSD.

\section{Statistics-based SPP estimation}\label{statistics-based}
In this section, we briefly review the baseline of the $a$ $posteriori$ SPP estimator proposed in \cite{gerkmann2011unbiased}, which is used to calculate the target from the clean speech in Section IV. With two hypotheses that $\mathcal{H}_0$ and $\mathcal{H}_1$ represent speech absence and presence, respectively, i.e.,
\begin{equation}
\left\{
\begin{aligned}
&\mathcal{H}_{0} :  Y(l)=N(l)\\
&\mathcal{H}_{1} :  Y(l)=X(l) + N(l).
\end{aligned}
\right.
\label{eq4}
\end{equation}

Subsequently, the likelihood function of the observed signal $Y(l)$ for each hypothesis can be obtained as
\begin{equation}
    p\left(Y(l)|\mathcal{H}_0\right)=\frac{1}{\pi\Phi_{N}(l)} e^{ -\frac{|Y(l)|^2}{\Phi_{N}(l)}},
    \label{eq5}
\end{equation}
\begin{equation}
\begin{split}
        p\left(Y(l)|\mathcal{H}_1\right) &=\frac{1}{\pi(\Phi_{N}(l)+\Phi_{X}(l))} e^{-\frac{|Y(l)|^2}{\Phi_{N}(l)+\Phi_{X}(l)}} \\
        &=\frac{1}{\pi\Phi_{N}(l)(1+\xi_{\mathcal{H}_{1}}(l))}e^{ -\frac{|Y(l)|^2}{\Phi_{N}(l) \left(1+\xi_{\mathcal{H}_{1}}(l)\right)}},
\end{split}
    \label{eq6}
\end{equation}
where $\xi_{H_1}(l)=\Phi_{X}(l)/\Phi_{N}(l)$ is the $a$ $priori$ SNR for speech presence. According to Bayes' theorem, the $a$ $posteriori$ SPP can be obtained by
\begin{equation}
    \begin{split}
        p(\mathcal{H}_1|Y(l))&=\frac{p(\mathcal{H}_1)p(Y(l)|\mathcal{H}_1)}{p(\mathcal{H}_0)p(Y(l)|\mathcal{H}_0) + p(\mathcal{H}_1)p(Y(l)|\mathcal{H}_1)} \\
        &=\left(1+ \alpha\left(1+\xi_{\mathcal{H}_1}(l) \right) e^{-\frac{|Y(l)|^2}{\Phi_{N}(l)}\frac{\xi_{\mathcal{H}_1}(l)}{1+\xi_{\mathcal{H}_1}(l)} }\right)^{-1},
    \end{split}
    \label{eq7}
\end{equation}
where $p(\mathcal{H}_{1}) = 1 - p(\mathcal{H}_{0})$ is the $a$ $priori$ SPP, $\alpha=p(\mathcal{H}_{0})/p(\mathcal{H}_{1})$. To simplify, $p(\mathcal{H}_{0})$ and $p(\mathcal{H}_{1})$ are commonly used as 0.5 to estimate $p(\mathcal{H}_1|Y(l))$ in \cite{gerkmann2008improved, gerkmann2011unbiased}. Additionally, since a fixed $a$ $priori$ SNR $\xi_{\mathcal{H}_{1}}$ enables (\ref{eq6}) to be close to zero when speech is absent \cite{gerkmann2008improved, gerkmann2011unbiased}, one fixed optimal $a$ $priori$ SNR is given by \cite{gerkmann2011unbiased} that is $10\log_{10}(\xi_{\mathcal{H}_1}(l))=15$ dB by minimizing the total probability error of \cite{van2004detection}. Finally, to avoid a stagnation of the noise PSD update based on SPP, $p(\mathcal{H}_1|Y(l))$ is recursively smoothed over time, which can be written as
\begin{equation}
    \hat{p}(l)=\beta\hat{p}(l-1)+(1-\beta)p(l),
    \label{eq8}
\end{equation}
where $0 < \beta < 1$ is the smoothing factor chosen as 0.9 in \cite{gerkmann2011unbiased}. Now, if $\hat{p}(\mathcal{H}_1|Y(l))$ is larger than $\lambda$, $p(\mathcal{H}_1|Y(l))$ is forced to be lower than $\lambda$, i.e.,
\begin{equation}
p(\mathcal{H}_1|Y(l))=\left\{
\begin{aligned}
&min(\lambda, p(\mathcal{H}_1|Y(l))) & , & \hat{p}(l) > \lambda\\
&p(\mathcal{H}_1|Y(l)) & , & else.
\end{aligned}
\right.
\label{eq9}
\end{equation}
where $\lambda = 0.99$ in \cite{gerkmann2011unbiased}.

\section{Proposed Learning-based SPP Estimation and applications}\label{learning-based model}
In this section, the learning-based $a$ $posteriori$ SPP estimation approach is proposed and its applications are also proposed. In the first part of this section, since the input features and learning target have a great impact on learning-based SPP estimation, we consider more appropriate input features and target that can promote the supervised learning-based SPP estimator, which has the benefit of decoupling the noise power estimator from the subsequent steps in a speech enhancement framework. Then, a DNN model architecture and the training strategy are presented. Finally, the learning-based SPP estimation is integrated with some applications, i.e., noise PSD estimation and speech enhancement in this paper.

\subsection{Input Features and Learning Target}\label{section4A}
In this paper, the supervised learning-based approach is used to estimate SPP. Therefore, we use the training data pairs to train a DNN model. Training data pairs consist of the input features and the learning target.

Firstly, since the statistics-based SPP estimator described in Section \ref{statistics-based} works in the STFT domain, the learning-based SPP estimation scheme also operates in the STFT domain. Considering that the log-power spectral density can offer perceptually relevant parameters \cite{xie1994family, wan1999networks, xu2014regression} which is conducive to speech signal processing such as noise reduction and speech enhancement in the STFT domain, therefore, to estimate SPP from the observed signal using the DNN, the log power spectral of the observed signal is used as the input feature in this paper.
It is defined as
\begin{equation}
    \Phi^{'}_Y(l)={\rm{log}}(\Phi_{Y}(l))
    \label{eq10}
\end{equation}

With the log-power spectral density of the observed signal as the input feature, the learning target of the $a$ $posteriori$ SPP should be provided. Although in Section \ref{statistics-based}, an optimal SPP can be provided as the learning target, we found that since the $a$ $priori$ SNR and SPP are fixed, it may degrade the performance of the learning-based $a$ $posteriori$ SPP estimator. To overcome these problems, we provide a more appropriate learning target for the learning-based SPP estimator.

While it is hard to estimate the $a$ $priori$ SPP when performing the $a$ $posteriori$ SPP estimation in (\ref{eq7}), the actual speech and noise power spectra are available during training. Therefore, the $a$ $priori$ SNR can be obtained directly from the training data. To make the $priori$ SPP independent from the observation, the uniform prior that is $P(\mathcal{H}_{0})=P(\mathcal{H}_{1})=0.5$ was adopted in \cite{gerkmann2011unbiased}. However, \cite{gerkmann2011unbiased} also mentioned that using the uniform prior is a worst-case assumption. Moreover, \cite{ephraim1985speech} demonstrated that the $a$ $priori$ SPP greatly impacts the $a$ $posteriori$ SPP estimation. Therefore, to improve the performance of the $a$ $posteriori$ SPP estimation, we suggest using the updated $a$ $priori$ SPP with the observation in contrast to the fixed prior in \cite{gerkmann2011unbiased}. 

In \cite{martin2005speech}, the optimal MMSE estimator is linear when the spectral coefficient of noise and speech are complex Gaussian distributions. The linear MMSE estimator is given by
\begin{equation}
\begin{split}
    E\left[X(l)|Y(l)\right]&= \frac{\Phi_{X}(l)}{\Phi_{X}(l) + \Phi_{N}(l)}Y(l)\\
    &=G(l)Y(l),
\end{split}
    \label{eq11}
\end{equation}
where $G(l)=\Phi_{X}(l)/(\Phi_{X}(l)+\Phi_{N}(l))$ is the gain function, and the value lies between 0 and 1. Therefore, the gain function is regarded as the soft gain for this linear estimator to decide speech presence probability in each T-F bin. Inspired by this, this paper employs the gain function to replace the fixed priors during training, i.e., $P(\mathcal{H}_{1})=G(l)$.

Subsequently, using the actual $a$ $priori$ SNR and the gain function in (\ref{eq11}) to replace the fixed  $a$ $priori$ SNR and SPP in (\ref{eq7}), it can be rewritten as
\begin{equation}
    p(\mathcal{H}_{1}|Y(l))=\left(1+ \left(1+\frac{1}{\xi_{\mathcal{H}_1}(l)} \right) e^{-\frac{|Y(l)|^2}{\Phi_{N}(l)}\frac{\xi_{\mathcal{H}_1}(l)}{1+\xi_{\mathcal{H}_1}(l)} }\right)^{-1}.
    \label{eq12}
\end{equation}
Although we update the $a$ $priori$ SPP and SNR during the $a$ $posteriori$ SPP estimation, in contrast to \cite{cohen2003noise, sohn1998voice} which also update the $a$ $priori$ SNR with the observation, our modified the $a$ $posteriori$ SPP is more reasonable because in \cite{cohen2003noise, sohn1998voice} when the $priori$ SNR $\xi_{\mathcal{H}_{1}}$ is zero, the $a$ $posteriori$ SPP yields only the prior, i.e., $p(\mathcal{H}_{1}|Y(l), \xi_{\mathcal{H}_{1}}=0)=p(\mathcal{H}_{1})$. Therefore, when employing the adaptive priors proposed in \cite{cohen2003noise, sohn1998voice}, and having a zero $a$ $priori$ SNR, the $a$ $posteriori$ SPP fails to converge to zero, indicating an inability to detect the absence of speech. With modifications in (\ref{eq12}), it also enables the $a$ $posteriori$ SPP to be close to zero when the updated $\xi_{\mathcal{H}_{1}}$ is zero same as \cite{gerkmann2011unbiased} that is speech absence can be detected.

\subsection{Model Architecture and Training Strategy}
With the input features and learning target in Section \ref{section4A}, (\ref{eq10}) and (\ref{eq12}) are employed as the training data pairs to train a DNN model in a supervised way. To improve the performance of the learning-based SPP estimator, we thoroughly analyzed the frequency bin-wise-based approach in \cite{tao2023frequency} and \cite{tao2023single}, which can not only reduce the model complexity sharply but also improve the performance of the $a$ $posteriori$ SPP estimation. In this paper, we aim to further improve the performance of the frequency bin-wise-based SPP estimator, which can be applied to noise estimation and speech enhancement with a high noise estimate accuracy and speech quality.

In \cite{tao2023frequency}, we decoupled frequency bins from the log-power spectral density, which can be described as
\begin{equation}
    \boldsymbol{\Phi}^{'}_{Y}(k)=\left[ \Phi^{'}_{Y}(k, 0),...,\Phi^{'}_{Y}(k,L-1) \right]^{T},
    \label{eq13}
\end{equation}
where $\boldsymbol{\Phi}^{'}_{Y}(k)$ is the vector of each frequency bin which contains $L$ time frames and $T$ denotes transpose operation. Each frequency contains $L$ T-F bins. Therefore, $K$ frequency bin vectors can be decoupled from the log-power spectral density. Additionally, the learning target in (\ref{eq7}) was also decoupled for each frequency bin vector. Using the decoupled training data pairs, $K$ DNNs were trained independently. To improve the performance of the SPP estimation, each frequency bin vector with its neighboring frequency bin vectors was employed as the input features to train the DNN, which can help the DNN learn more about the representation from the observed signal to better estimate SPP. Although using the decoupled training data pairs to train multiple DNNs can reduce model complexity, the performance of the SPP estimation would be degraded, especially in adverse noise environments, because each DNN only focuses on local information.

To extend the possibilities of the frequency bin-wise-based SPP estimation approach, in \cite{tao2023frequency}, taking global information into account, we introduced the global information of the input features to the decoupled frequency bin vectors to generate the hybrid global-local information. In contrast to the high dimensional input features, the global information is represented by a low dimensional latent space, which is extracted from the input features using an encoder. The results shown in \cite{tao2023single} demonstrated the importance of the global information for SPP estimation because it can significantly improve the performance of SPP estimation, especially in low SNR conditions.

\begin{figure}[ht]
  \centering
  \includegraphics[width=0.4\linewidth]{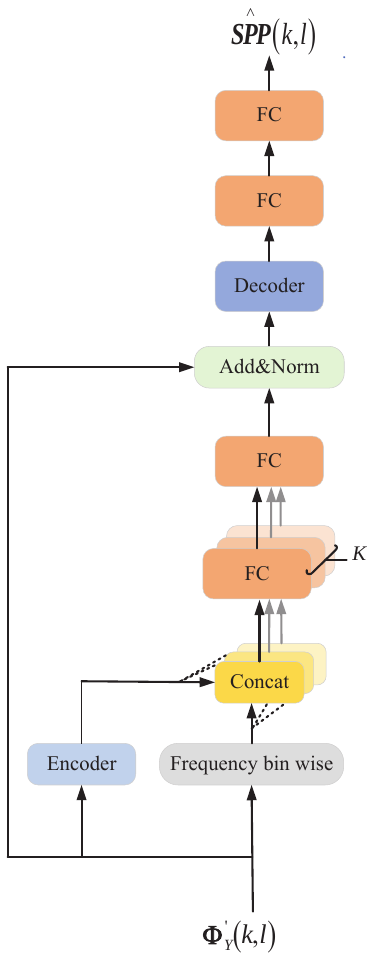}
  \caption{The Learning-based SPP estimation framework. It consists of an alternative encoder and decoder, and the fully connected layer (FC).}
  \label{SPP Model}
\end{figure}

In this paper, based on our previous work, local and global information are also used to train a high-performance SPP decoder to estimate SPP. Firstly, each frequency bin vector $\bm{\Phi}_{Y}^{'}(k)$ is decoupled from the $\Phi^{'}_{Y}(l)$, and an alternative encoder is used to transform global information from the high-dimensional input features of $\Phi^{'}_{Y}(l)$ into a low-dimensional latent space. Then, each frequency bin vector independently concatenates the latent space as the input of $K$ FC layers. The $K$ outputs are then used as the input of an FC layer, and there is a residual connection between the output of the second FC layer and $\bm{\Phi}_{Y}^{'}(k)$. Finally, the residual connection output is used to train an alternative decoder and two FC layers. Fig. (\ref{SPP Model}) shows the learning-based SPP estimation framework. The details of the model composition are described in Section \ref{model composition}.

As the input of the SPP estimation model, it consists of all frequency bins. Therefore, the input matrix $\bm{\Phi}^{'}_{Y}(k,l)$ can be obtained as 
\begin{equation}
\boldsymbol{\Phi}^{'}_{Y}(k,l)=
\begin{bmatrix}
\Phi^{'}_{Y}(0,0) & \dots & \Phi^{'}_{Y}(0,L-1) \\
\vdots & \ddots & \vdots \\
\Phi^{'}_{Y}(K-1,0)& \dots & \Phi^{'}_{Y}(K-1,L-1) \\
\end{bmatrix}.
\label{eq14}
\end{equation}

Then, using the SPP estimator shown in Fig. \ref{SPP Model} to estimate the $a$ $posteriori$ SPP for each time-frequency bin, we have
\begin{equation}
    \widehat{\boldsymbol{SPP}}(k,l)=F^{\Theta} \{\boldsymbol{\Phi}_{Y}^{'}(k,l) \},
    \label{eq15}
\end{equation}
where $\widehat{\boldsymbol{SPP}}(k,l)$ is SPP estimate, $F^{\Theta}$ is the DNN model with the parameter $\Theta$. During training, the Kullback-Leibler \cite{hershey2007approximating} divergence is employed as the loss function to calculate the training loss, i.e.,
\begin{small}
\begin{equation}
    \mathcal{L}(\boldsymbol{SPP}(k,l), \widehat{\boldsymbol{SPP}}(k,l)) = {\boldsymbol{SPP}(k,l){\rm{log}}\left(\frac{\boldsymbol{SPP}(k,l)}{\widehat{\boldsymbol{SPP}}(k,l)}\right)}.
    \label{eq16}
\end{equation}
where $\mathcal{L}(\boldsymbol{SPP}(k,l), \widehat{\boldsymbol{SPP}}(k,l))$ is the loss error, and $\boldsymbol{SPP}(k,l)$ is the learning target derived by (\ref{eq7}).
\begin{equation}
\boldsymbol{SPP}(k,l)=
\begin{bmatrix}
p(0,0) & \dots & p(0,L-1) \\
\vdots & \ddots & \vdots \\
p(K-1,0)& \dots & p(K-1,Y-1) \\
\end{bmatrix},
\label{eq17}
\end{equation}    
\end{small}

where $p(k,l)$ is the $a$ $posteriori$ SPP for the T-F bin whose frame index is $k$ and frequency index is $l$. Finally, Adam \cite{kingma2014adam} is employed as the optimizer to update the parameter $\Theta$ during training.

\subsection{SPP-based Noise PSD Estimation and Speech Enhancement with Sub-Optimal MMSE}
\subsubsection{Noise PSD estimation}
For the statistics-based SPP estimator shown in Section \ref{statistics-based}, the Unbiased MMSE-based noise PSD estimation proposed in \cite{gerkmann2011unbiased} is given by
\begin{equation}
    E\left[ |N(l)|^{2}|Y(l)\right]=p(\mathcal{H}_{0}|Y(l))|Y(l)|^{2} + p(\mathcal{H}_{1}|Y(l))\hat{\Phi}_{N}(l),
    \label{eq18}
\end{equation}
where $\hat{\Phi}_{N}(l)$ is the noise PSD estimate. And in \cite{gerkmann2011unbiased}, $\hat{\Phi}_{N}(l)$ denotes the previous frame noise PSD estimate.

Compared to the MMSE-based noise PSD estimator in \cite{hendriks2010mmse}, (\ref{eq18}) employed a soft decision SPP to replace the hard decision of the VAD. Therefore the bias compensation is unnecessary for the SPP-based noise PSD estimation in (\ref{eq18}). Then the noise PSD is smoothed recursively, i.e.,
\begin{equation}
    \hat{\Phi}_{N}(l)=c\hat{\Phi}_{N}(l-1)+(1-c)E\left[ |N(l)|^{2}|Y(l)\right],
    \label{eq19}
\end{equation}
where $0 < c < 1$ is a smoothing factor.

In (\ref{eq18}) and (\ref{eq19}), the previous frame noise PSD estimate is used to estimate the current frame noise PSD. Therefore, estimating the current frame noise PSD depends on the noise PSD estimate of the previous frame. In \cite{gerkmann2011noise}, the $a$ $posteriori$ SPP estimator also uses the estimated noise PSD of the previous frame to estimate the current frame $a$ $posteriori$ SPP in (\ref{eq7}). For this reason, the previous frame SPP estimate and noise PSD estimate significantly impact the current frame SPP estimation accuracy. To reduce the impact of the previous estimate on the current frame, we proposed the learning target of the learning-based SPP estimator in (\ref{eq12}), which is computed using the actual noise PSD of the current frame. Furthermore, since the recurrent neural network is used to design the SPP estimator, our proposed SPP estimator can acquire contextual information from a few input frames to estimate the $a$ $posteriori$ SPP of the current frame without the estimated noise PSD of the previous frame. Consequently, in this paper, the sub-optimal MMSE is used to estimate noise PSD based on the estimated SPP. The sub-optimal MMSE noise estimator can be obtained as
\begin{equation}
    E\left[ |N(l)|^{2}|Y(l)\right]=p(\mathcal{H}_{0}|Y(l))|Y(l)|^{2}.
    \label{eq20}
\end{equation}

\subsubsection{Speech Enhancement}
With the SPP estimate, the optimal noise statistics are obtained which can further facilitate the usage of a wide range of optimal filtering techniques to optimally control the trade-off between noise reduction and signal distortion \cite{nielsen2023analysis}. In this paper, we employ a standard speech enhancement framework to enhance speech from the observed noisy speech. Since the log spectral amplitude (LSA) estimator \cite{ephraim1985speech} is robust against the error of noise PSD estimation, which can reduce musical noise, the estimated noise PSD is incorporated into LSA to enhance speech. In \cite{ephraim1985speech}, the optimal LSA estimator is obtained as
\begin{equation}
    \hat{X}(l)=\frac{\xi(l)}{\xi(l) + 1}e^{\frac{1}{2}\int^{\infty}_{v(l)}\frac{e^{-t}}{t}dt}Y(l)
    \label{eq21}
\end{equation}
and
\begin{equation}
    v(l)=\frac{\xi(l)}{1+\xi(l)}\gamma(l)
\end{equation}

where $\xi(l)$ is the $a$ $priori$ SNR, $\gamma(l)$ is the $a$ $posteriori$ SNR, and $\hat{X}(l)$ is the enhanced speech. In \cite{ephraim1985speech}, a limited maximum-likelihood (ML) estimate is employed to estimate the $a$ $priori$ SNR. It can be obtained as
\begin{equation}
    \hat{\xi}(l)=\rm{max}(0, \hat{\xi}^{ml}(l)) = \rm{max}(0, \hat{\gamma}(l) - 1)
    \label{eq22}
\end{equation}
where $\hat{\xi}(l)$ is the estimated $a$ $priori$ SNR, $\hat{\xi}^{ml}(l)$ is the ML estimated $a$ $priori$ SNR, $\hat{\gamma}(l)$ is the estimated $a$ $posteriori$ SNR. Although the speech enhancement performance of the LSA estimator mostly depends on the noise PSD estimation accuracy, to reduce the impact of noise estimation error on speech enhancement, the \textit{decision-directed} approach \cite{ephraim1985speech, cohen2003noise, gerkmann2008improved} is employed to update the $a$ $priori$ SNR estimate within the LSA estimator. The $a$ $priori$ SNR is given as
\begin{equation}
    \hat{\xi}(l)=\alpha_{snr}\frac{|Y(l-1)|^{2}}{\Phi_{N}(l-1)}+(1-\alpha_{snr})(\hat{\gamma}(l)-1),
    \label{eq23}
\end{equation}
where $\alpha_{snr}$ is the important tuning factor for the $a$ $priori$ SNR estimate. The lower bound of $\hat{\xi}(l)$ is set to -25 dB.

With the estimated noise PSD in (\ref{eq20}), we can obtain
\begin{equation}
    \hat{\gamma}(l) = \frac{\Phi_{Y}(l)}{\hat{\Phi}_{N}(l)}
    \label{eq24}
\end{equation}

Therefore, with the $a$ $posteriori$ SPP estimate, the noise PSD can be estimated by using (\ref{eq20}), and then the estimated noise SPP is employed to enhance speech with (\ref{eq21})-(\ref{eq24}). 

\section{Implementation Details}\label{implementation}
In this section, we provide detailed information on experimental settings and evaluation metrics. Additionally, to further prove the effectiveness of the proposed method, the proposed SPP estimator is deployed into noise estimation and speech enhancement. Since noise estimation and speech enhancement are based on SPP estimation, the performance of noise estimation and speech enhancement depends on the performance of SPP estimation. Therefore, the evaluation of noise estimation and speech enhancement are also provided. Firstly, the dataset is presented which is used to train, validate, and test. Then, the detail of the model composition for our proposed learning-based SPP estimator is provided. Finally, the evaluation metrics of SPP estimation, noise estimation, and speech enhancement are presented.

\subsection{Experimental Settings}
\subsubsection{Dataset}
The Deep Noise Suppression (DNS) challenge dataset \cite{reddy2021icassp} is used to train, validate, and test the proposed learning-based $a$ $posteriori$ SPP estimator in Section \ref{learning-based model}. All clean and noise utterances are recorded by a single-channel microphone with a 16 kHz sampling frequency. For training, 48600 clean utterances are randomly corrupted by 48600 noise utterances to generate 48600 noisy utterances. When we generate noisy utterances, the signal-noise ratio (SNR) varies between -10 dB to 10 dB, in steps of 1 dB. 7200 utterances also with SNR varying between -10 dB to 10 dB in 1 dB increments as the validation set. For training and validation, each utterance is 2 seconds. More importantly, when we generate the datasets, we use different clean and noise utterances to generate noisy utterances, which can guarantee the datasets are different. After training, 0.5 hours of utterance with random SNR from -10 dB to 10 dB is generated using the DNS dataset to test the SPP estimation accuracy and the downstream task performance. For testing in the DNS dataset, each utterance is 20s. Additionally, to test the performance of SPP-based noise PSD estimation and speech enhancement, the unseen noise datasets NOISEX-92 \cite{varga1993assessment}, which contains 15 different noise types, are resampled from 19.98 kHz to 16 kHz to generate the testing set. 1800 clean utterances from the DNS dataset are randomly corrupted by the noise from NOISEX-92 to generate 1800 noisy utterances, and each utterance is 2 seconds. For testing, we chose the testing set for five different levels of the input SNR, which are -10 dB, -5 dB, 0 dB, 5 dB, and 10 dB. Under different SNR levels, there are 90 utterances (0.5 hours) from DNS datasets and 1800 utterances (1 hour) generated by DNS clean utterances and NOISEX-92 noise utterances.

\subsubsection{Pre Processing}
Before using the data to train the learning-based SPP estimation model, we pre-process the dataset that can meet the requirement of the model input. Firstly, we use the mean and standard derivation to normalize the dataset. Then, Hamming window is employed in the STFT analysis. The window length is 16ms, and the overlap length is 8ms.
\subsubsection{Training Procedure}
During training, the learning rate is set to 0.001. The weight decay is set to 0.00001 to prevent overfitting. The mini-batch size is 64 with a shuffle. Our proposed and other reference DNN-based SPP estimators are trained within 100 epochs. Subsequently, we choose the trained model according to the performance on the validation set. To save training time, a patience factor is set to 10 epochs that is the training will be stopped when the validation loss can not decrease more than 10 epochs.
\subsection{Model composition}\label{model composition}
For our proposed learning-based SPP estimator shown in Fig. \ref{SPP Model}, an alternative encoder and decoder, and fully connected (FC) layers are used to construct the model, which can process the speech signal with variable time length. Firstly, one encoder with 129 input nodes and 32 output nodes is used to extract global information from input features. With the FC layer output, 129 frequency bin vectors are concatenated with it as the input of 129 independent FC layers, which have 33 input nodes and one output node. All outputs of FC layers are connected with the log-power spectrogram with layer normalization as the input of the SPP decoder. The decoder is an alternative that can process the time series. In this work, to compare with two baselines \cite{tammen2020dnn} and \cite{xia2020weighted}, we use two different encoders and decoders to achieve high performance with a reasonable model complexity. (1) Since \cite{tammen2020dnn} used one bidirectional long short-term memory (BLSTM) to consist the DNN model, we employ one BLSTM as the decoder, and to achieve a better performance one LSTM is used as the encoder. (2) To achieve real-time processing capability, the decoder is two multi-head attention layers with 3 heads and the encoder is one FC layer. The input and output dimensions of the decoder are 129 and 129, respectively. After the decoder, one FC layer with 258 input nodes and 258 output nodes, and one FC layer with 258 input nodes and 129 output nodes are placed. Finally, the output layer is the activation function Sigmoid to restrict the output range from 0 to 1. To clarify our proposed DNN model in the following experiments, we define the models (1) and (2) are Proposed (BLSTM) and Proposed (Attention), respectively.

\subsection{Evaluation Metrics}
\subsubsection{SPP estimation}
To evaluate the SPP estimation performance, one case study for the distribution of the SPP on the time-frequency plane is shown. We randomly select an utterance corrupted by the additive noise from the testing set. Additionally, the receiver operating characteristic (ROC) curve is also computed to evaluate speech detection accuracy when interpreting the SPP estimator as the detector. To compute the ROC curves, we pre-defined a threshold to distinguish speech components and non-speech components among the ground truth which is derived from (\ref{eq12}) with actual speech and noise. Finally, according to the ROC curve, the area under the curve (AUC) and speech detection accuracy ($P_{d}$) are computed. For speech detection accuracy, the true positive rate is regarded as the speech detection accuracy when the false alarm rate ($P_{fa}$) is 0.05 \cite{momeni2014single, chang2006voice}.

\subsubsection{Noise PSD estimation accuracy}
With the estimated noise PSD $\hat{\Phi}_{N}(l)$ and the reference noise PSD $\Phi_{N}(l)$, we employ the symmetric mean log-spectral distortion (LogErr) \cite{hendriks2008noise} to measure the accuracy between $\hat{\Phi}_{N}(l)$ and $\Phi_{N}(l)$. It can be described as
\begin{equation}
    {\rm{LogErr}} = \frac{1}{KL}\sum^{K-1}_{k=0}\sum^{L-1}_{l=0}\left| 10{\rm{log}}_{10} \left[ \frac{\Phi_{N}(k,l)}{\hat{\Phi}_{N}(k,l)}\right] \right|
    \label{eq25}
\end{equation}
\subsubsection{Speech enhancement performance}
For the simulated data from DNS dataset, the objective evaluator DNSMOS is employed to compute the speech distortion (SIG) and residual noise (BAK). Additionally, the objective measurement is used to evaluate enhanced speech quality and intelligibility. Since the perceptual evaluation of speech quality (PESQ)  \cite{recommendation2001perceptual} is primarily determined by the type and degree of distortion, speech content, and cognitive modeling, the wide-band PESQ whose score ranges from -0.5 to 4.5 is employed to evaluate the objective speech quality. Given the clean speech, the PESQ score of the enhanced speech is computed and a higher PESQ score denotes better speech quality. Based on the correlation coefficient between the temporal envelops of the clean speech and the enhanced speech, the shot-time objective intelligibility (STOI) \cite{taal2011algorithm} is computed. STOI ranges from 0 to 1, representing the percent correct (0 to 100\%) and better speech intelligibility with a higher STOI score. Furthermore, for multi-channel speech enhancement performance evaluation, the composite measures for signal distortion (CSIG), and residual noise (CBAK) \cite{hu2007evaluation} scores are also computed.

\subsubsection{Computational complexity}
Since the SPP estimator could be deployed to speech enhancement work and our proposed method is learning-based, the number of parameters (Params) and floating point operations (FLOPs) are computed to represent the computational complexity. In this work, the Python library $ptflops$ \cite{ptflops} is used to compute the Params and FLOPs.

\begin{figure*}[ht]
\centering
	\subfloat[Clean Speech]{\includegraphics[width = 0.24\textwidth]{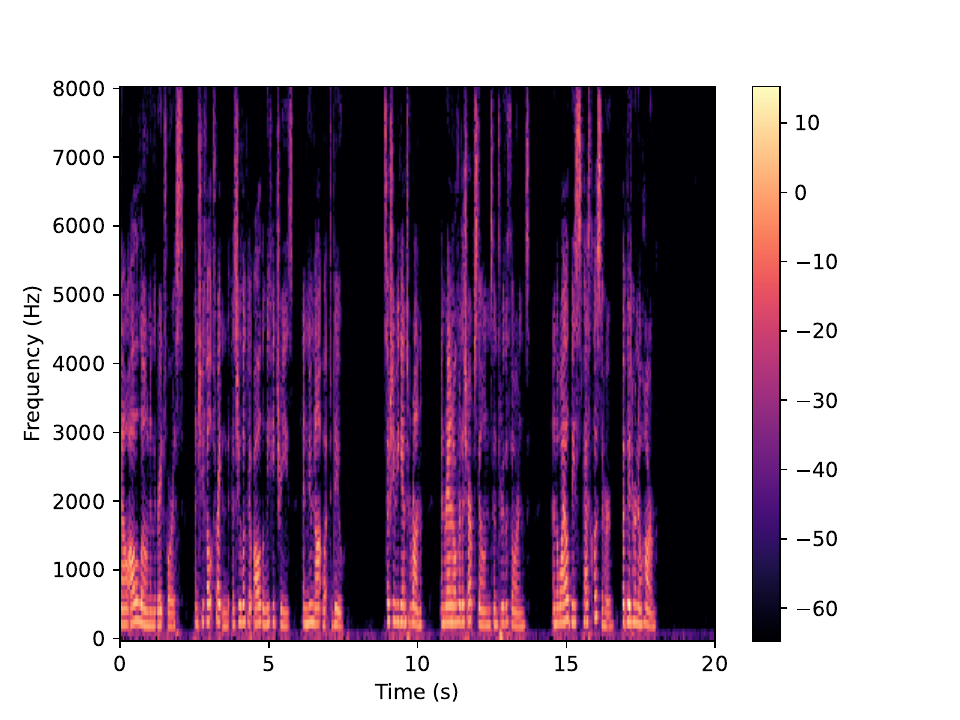}}
	\hfill
	\subfloat[Noisy Speech]{\includegraphics[width = 0.24\textwidth]{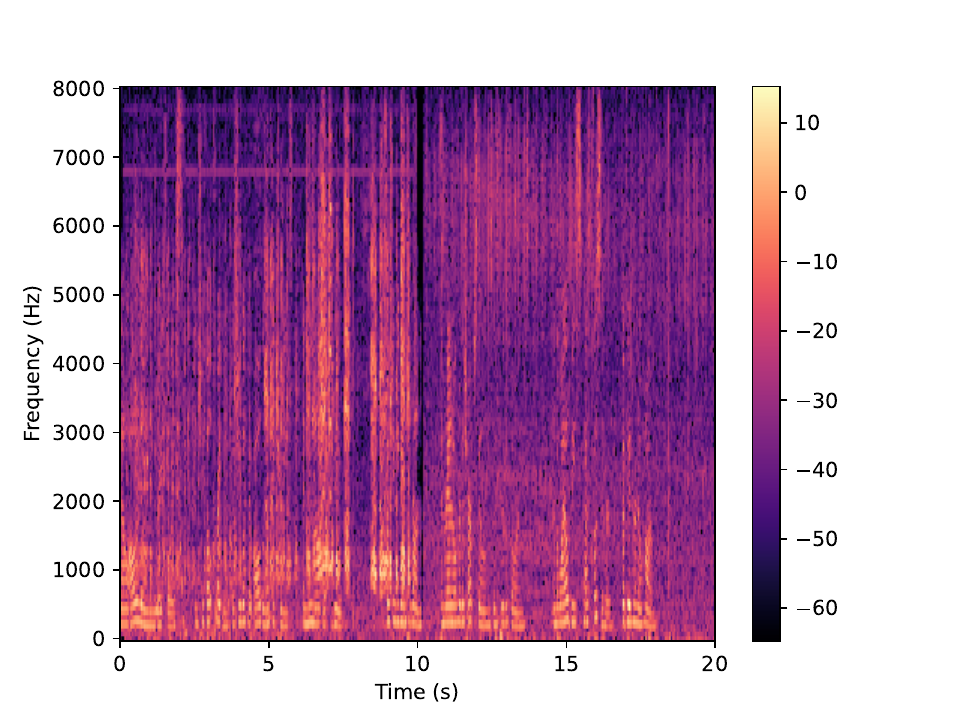}}
        \hfill
        \subfloat[Ground Truth (Proposed)]{\includegraphics[width = 0.24\textwidth]{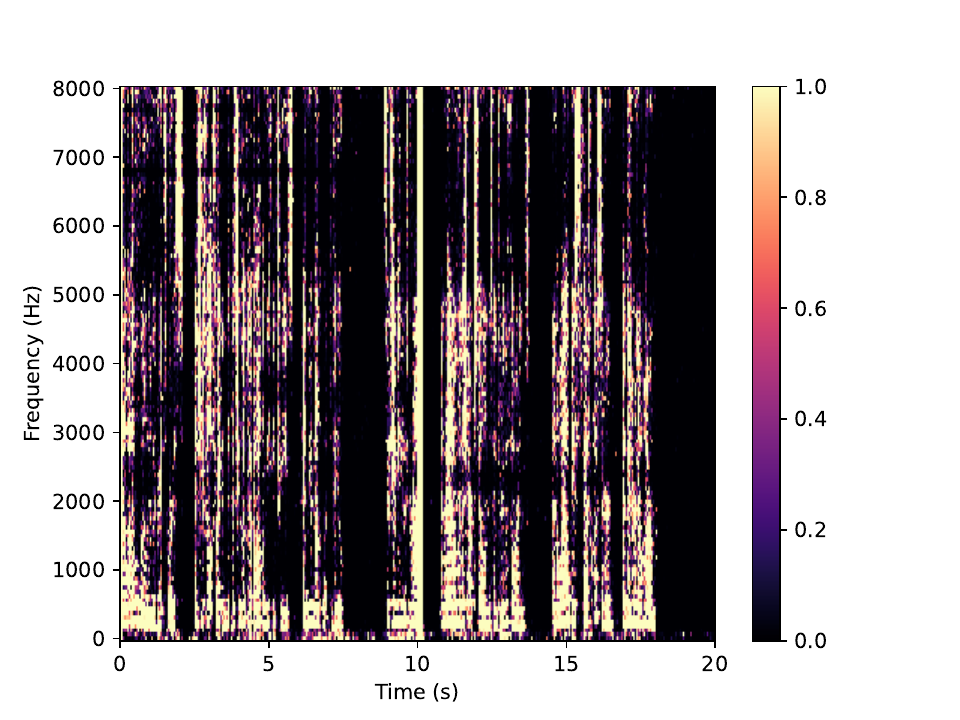}}
	\hfill
	\subfloat[Estimated SPP (Proposed)]{\includegraphics[width = 0.24\textwidth]{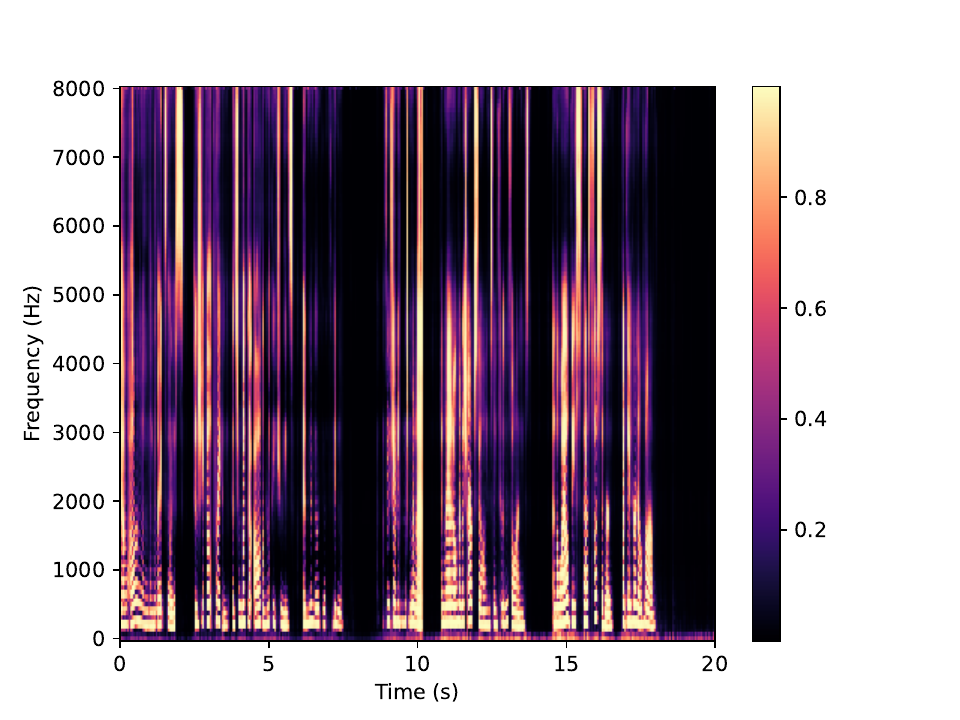}} 
        
\caption{In the STFT domain, the clean speech, the observed signal, and the distribution of the SPP are shown in (a), (b), and (c)-(d). (c) is the ground truth for SPP which is computed from the noisy speech with the actual noise. (\ref{eq12}) is employed to compute (c). (d) is the estimated SPP using our proposed method and with the BLSTM as the decoder. Input SNR = 5 dB.}
\label{SPP_EST}
\end{figure*}

\section{Results and Discussions}\label{resutls}
In this section, a set of experiments is performed to evaluate the performance of the SPP learning-based noise PSD estimation and speech enhancement. To compare, some existing SPP-based noise estimation methods are performed. Additionally, the ablation study is also performed to SPP estimation, noise PSD estimation, and speech enhancement when the hybrid global-local information is removed directly from the proposed DNN-based SPP estimator. Finally, both the single-channel speech enhancement and multi-channel speech tasks are performed based on the SPP estimator with the LSA estimator. The model complexity of the SPP-based speech enhancement system is also provided.

\subsection{Performance for the learning-based SPP estimation}\label{SPP performance}
One utterance in which SNR is 5 dB is used for testing and the duration of the utterance is 20 seconds. With the actual noise and clean speech, the ground truth from the observed signal is computed by (\ref{eq12}). Fig. \ref{SPP_EST} (c) and (d) show the distribution of the ground truth and the SPP estimate on the time-frequency plane, respectively. Compared to the clean speech and the observed signal shown in Fig. \ref{SPP_EST} (a) and (b), the ground truth can track the noise with the small value of the SPP estimate when the noise and speech exist simultaneously. More importantly, the speech interval can be detected accurately and the lowest value is very close to 0. For the SPP estimate computed by our proposed SPP estimator, we can find that it is very close to the ground truth and it can also detect the weak speech component, especially for the high-frequency area.

Additionally, we computed the ROC curve between the SPP estimate and the ground truth. To demonstrate the improvement of our proposed SPP estimator in this paper, we compared the SPP estimator proposed in \cite{tao2023single}. Although in \cite{tao2023single} we have demonstrated that using the hybrid global-local information can not only reduce the model complexity but also improve the SPP estimation accuracy, we extend the model by using a BLSTM/Attention layer as the decoder to further improve SPP estimation accuracy. One DNN-based SPP estimator \cite{tammen2020dnn}, which also consists of the BLSTM layer is implemented. For comparability, all SPP estimators are trained by the proposed learning target. For testing, 1 hour utterance is generated from the DNS challenge dataset. When generating the testing utterance, the input SNR is randomly selected from between -10 dB to 10 dB.

\begin{figure}
    \centering
    \includegraphics[width=\linewidth]{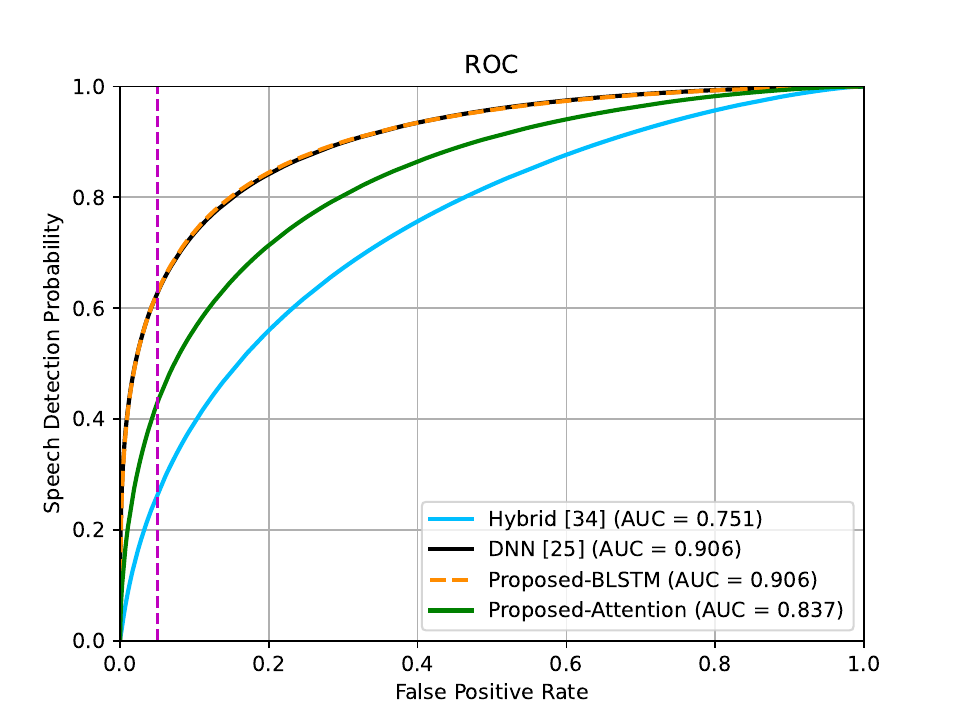}
    \caption{ROC Curves comparison for SPP estimation. Four different SPP estimators are trained by our proposed learning target. The predefined threshold is 0.135, and the purple dashed line is the false alarm rate of 0.05.}
    \label{roc}
\end{figure}

The predefined threshold in \cite{tao2023single} is also used to compute ROC curves in this paper. With experiments, we found that although the threshold can influence the AUC score and $P_{d}$ score, the variance in the score is very small. Additionally, when we use the same threshold to compute the ROC curves for different SPP estimators, the difference in the scores between different SPP estimators still changes minimally. Therefore, threshold selection does not affect the result of the ROC curve comparison.

In Fig. \ref{roc}, four different ROC curves are shown for the performance of SPP estimation. In terms of AUC and speech detection accuracy, our proposed method and the DNN-based method achieved high performance in SPP estimation. Although the hybrid method is lower than the other three methods, it is still effective because it got a high AUC score of 0.794. More importantly, since we emphasize the real-time deployment in \cite{tao2023single} and also in this paper, the model complexity is a significant performance indicator. Compared with \cite{tao2023single}, as we add one BLSTM/Attention layer in this work, the model complexity will be increased. However, since we also use the frequency bin-wised approach proposed in \cite{tao2023frequency}, multiple DNNs instead of the fully connected layer are used to compose the model, which helps to reduce the model complexity. Additionally, by comparing the three methods we proposed, i.e., Hybrid in \cite{tao2023single}, proposed (BLSTM), and proposed (Attention), we can confirm that improving the decoding capacity of the hybrid global-local information can further improve the accuracy of the SPP estimation. For a more comprehensive evaluation, we employ our proposed SPP estimator to noise PSD estimation in Section \ref{noise_est} and speech enhancement tasks in Section \ref{SE}. Moreover, to assess computational complexity, a comparison of model complexity is presented in Section \ref{complexity}.

\subsection{Performance of SPP-based noise PSD estimation}\label{noise_est}
\begin{figure}[ht]
    \centering
    \subfloat[DNS dataset]{\includegraphics[width = 0.45\textwidth]{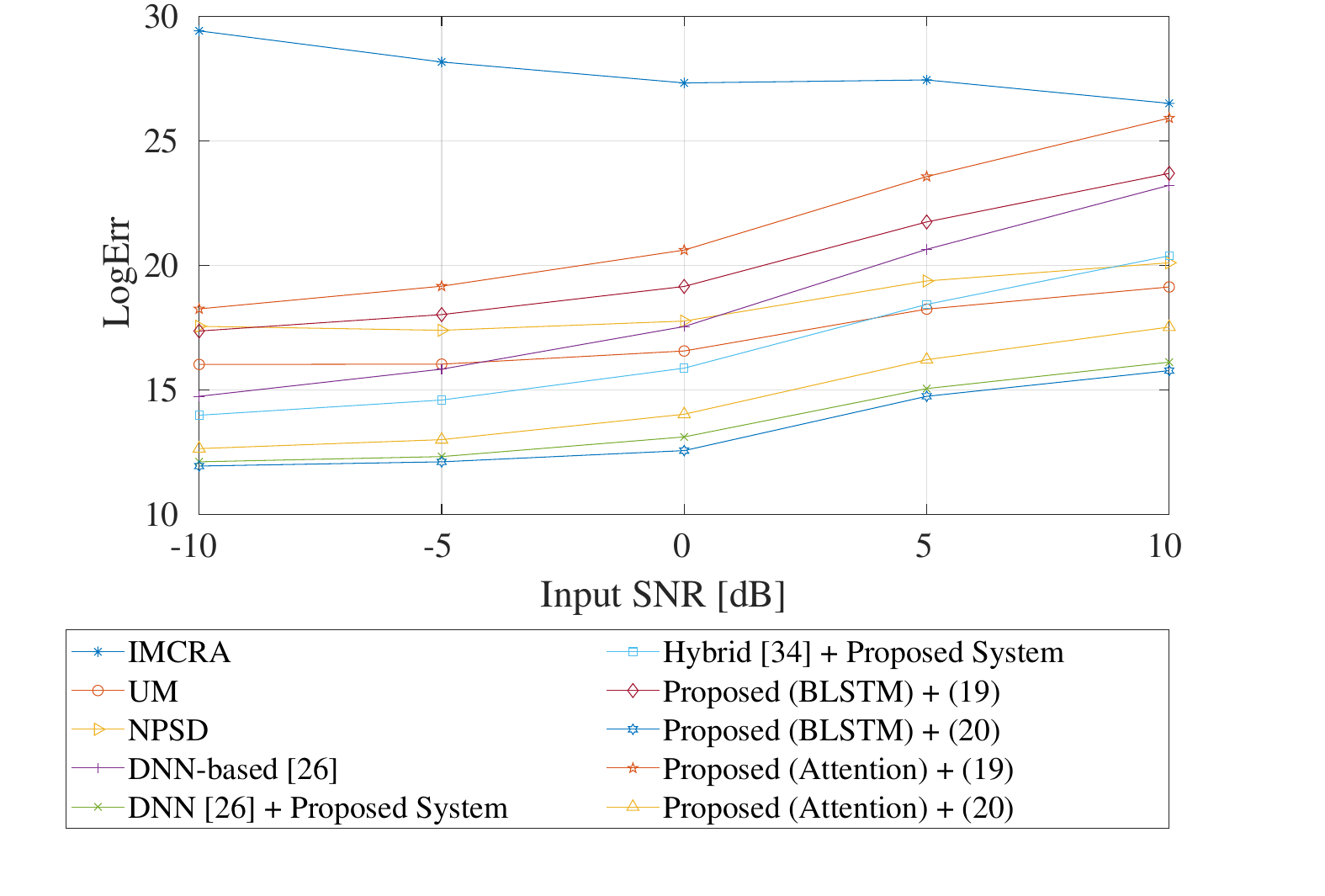}}\\
    \subfloat[NoiseX-92 dataset]{\includegraphics[width = 0.45\textwidth]{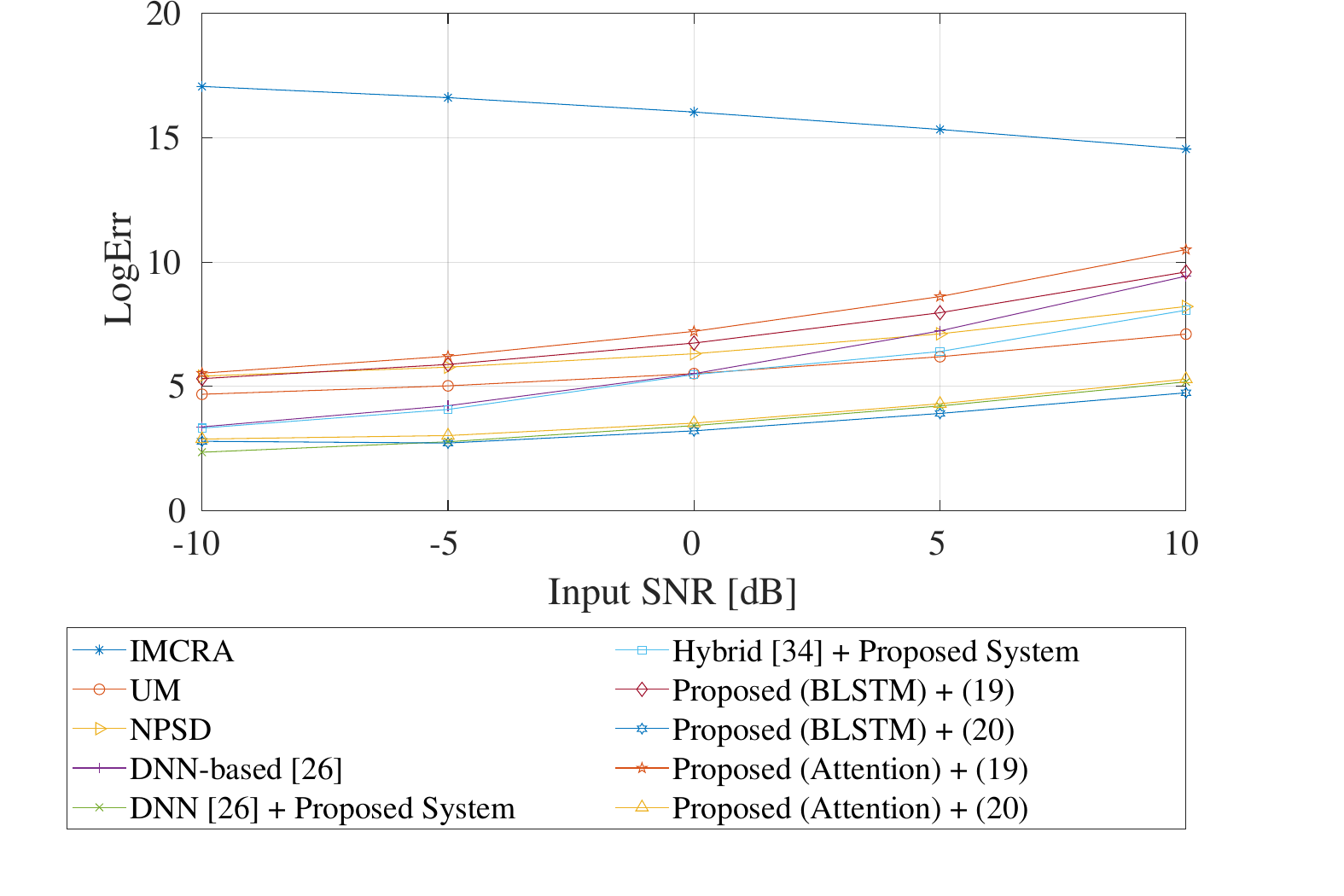}}
    \caption{The logarithmic errors of the various noise PSD estimators in \cite{cohen2003noise, gerkmann2011unbiased, li2016non, tammen2020dnn} and our proposed method which are based on the speech presence probability estimate. For the DNN-based approach in \cite{tammen2020dnn}, the deep neural network is applied to estimate SPP and then estimate noise PSD by using our proposed systems that are our proposed learning target and the sub-optimal MMSE-based noise PSD estimation approach. For our proposed SPP estimator, we use the optimal MMSE (\ref{eq19}) and sub-optimal MMSE (\ref{eq20}) to estimate noise PSD, respectively.}
    \label{LogErr}
\end{figure}

To compare the performance of the noise PSD estimation, the improved minima controlled recursive averaging (IMCRA) \cite{cohen2003noise}, unbiased MMSE (UM), and a non-stationary noise PSD estimator (NPSD) \cite{li2016non}, which are based on the SPP estimate and the optimal MMSE shown in (\ref{eq19}), are implemented. One deep learning-based noise PSD estimation approach which also employed a DNN to estimate SPP in \cite{tammen2020dnn} is implemented. In \cite{tammen2020dnn}, a different learning target for SPP estimation and the noise PSD estimation approach with a recursive way are used to enhance speech. To demonstrate the efficiency of our proposed learning target and noise PSD estimation approach, the DNN in \cite{tammen2020dnn} is trained by our proposed training strategy, and then the proposed noise PSD estimation approach in (\ref{eq20}) is used to estimate noise PSD. Additionally, to show the difference between the sub-optimal and optimal MMSE-based noise PSD estimation for our proposed learning-based SPP estimation method, we employ the optimal and sub-optimal MMSE to estimate noise PSD based on the estimated SPP. The LogErr is computed using (\ref{eq25}), and the results are given in Fig. \ref{LogErr}.

Among ten SPP-based noise PSD estimation approaches, the lowest LogErr scores were attained by two learning-based SPP estimators incorporating our proposed sub-optimal MMSE-based noise PSD estimation method. Therefore, we can confirm that our proposed learning-based SPP estimator is competent in tracking noise while it works with the sub-optimal MMSE. Additionally, for various noise datasets and input SNR levels, our proposed method with (\ref{eq20}) can still keep high performance, which can further demonstrate the capability and superiority of noise PSD tracking. For the DNN-based noise PSD estimator \cite{tammen2020dnn}, which also used the DNN to estimate SPP, we can find that it can not be interpreted as a noise PSD tracker as its LogErr is even higher than most conventional methods. While \cite{tammen2020dnn} utilized a high-performance DNN for SPP estimation, the inappropriate SPP learning target and flawed noise PSD estimation procedure lead to a deterioration in noise PSD estimation accuracy, rendering it less effective for noise PSD tracking. Moreover, when we employed our proposed training strategy to train the DNN in \cite{tammen2020dnn} and used (\ref{eq20}) to estimate noise PSD, we can find that the noise PSD estimation accuracy was improved compared to the noise PSD estimator in \cite{tammen2020dnn}. Therefore, we can observe that our proposed learning target in (\ref{eq12}) and noise PSD estimation approach in (\ref{eq20}) are effective for the learning-based SPP estimation directed noise PSD estimation.

\subsection{Performance of Single-Channel Speech Enhancement}\label{SE}
\begin{figure}[ht]
    \centering
    \subfloat[SIG]{\includegraphics[width = 0.5\textwidth]{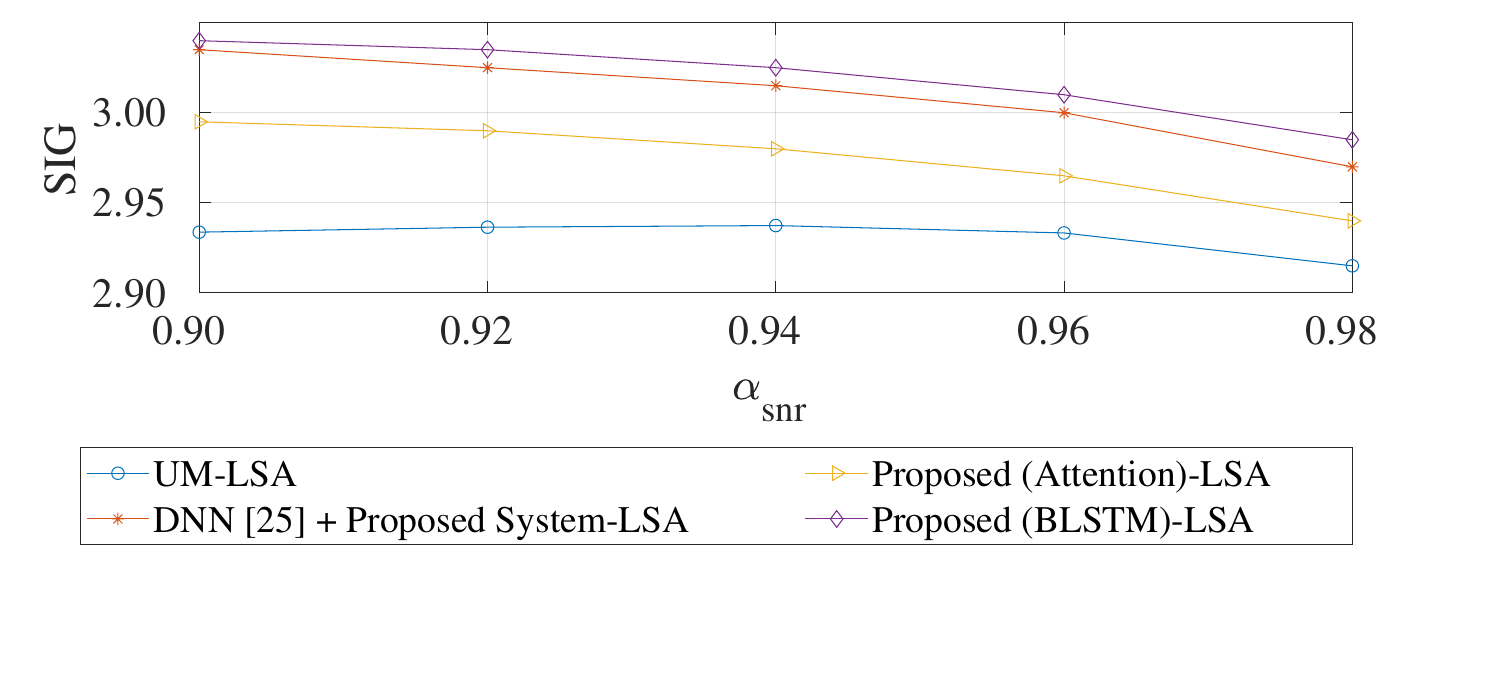}}\\
    \subfloat[BAK]{\includegraphics[width = 0.5\textwidth]{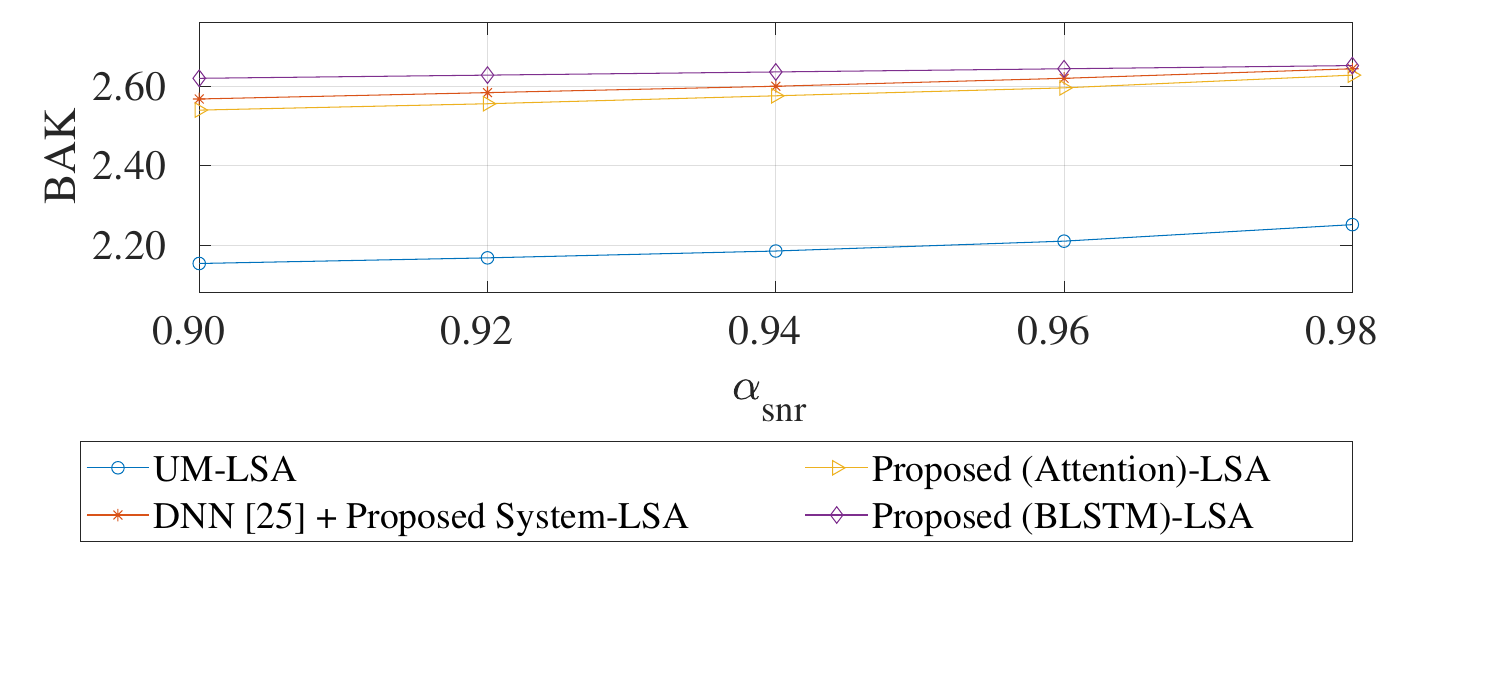}}\\
    \caption{SIG and BAK scores for the SPP-based speech enhancement evaluation in terms of speech distortion and residual noise. The scores are computed by the objective evaluator DNSMOS. For the speech enhancement framework, the UM and the learning-based SPP estimators (DNN in \cite{tammen2020dnn} and ours) are integrated with the LSA estimator. The learning-based SPP estimator (DNN in \cite{tammen2020dnn} and ours) uses our proposed sub-optimal MMSE to estimate noise.}
    \label{dnsmos}
\end{figure}

To extract clean speech from the observed signal, the noise PSD estimator proposed in Section \ref{noise_est} is employed to estimate the gain function combined with the LSA estimator. 

\begin{table*}[t]
  \centering
  \caption{PESQ score and STOI score for five different SPP-based-LSA speech enhancement frameworks and one deep learning-based speech enhancement method. Two different noise datasets, DNS and Noise-X92, are tested under various input SNR levels. The Proposed system is our proposed learning target and sub-optimal MMSE-based noise PSD estimation. The tunning factor of the LSA estimator is 0.90.\\}
  \scriptsize
  \begin{tabular}{*{14}{c}}
  \toprule
   \multirow{1}*{} &\multirow{3}*{Methods} & \multicolumn{6}{c}{PESQ} & \multicolumn{6}{c}{STOI} \\
   \cmidrule(lr){3-8}
   \cmidrule(lr){9-14}
   & & \multicolumn{5}{c}{Input SNR [dB]} & & \multicolumn{5}{c}{Input SNR [dB]} \\
   \cmidrule(lr){3-7}
   \cmidrule(lr){9-13}
  & &-10 & -5 & 0 & 5 & 10 & Average &-10 & -5 & 0 & 5 & 10 & Average\\
  \midrule
   \multirow{7}*{\rotatebox{90}{DNS}}
    &Noisy &1.34 &1.54 &1.79 &2.12 &2.43 & 1.84 &0.66 &0.74 &0.80 &0.86 &0.91 &0.79 \\
    &UM-LSA \cite{gerkmann2011unbiased} &1.29 &1.53 &1.87 &2.24 &2.58 &1.90 &0.63 &0.71 &0.78 &0.85 &0.90 & 0.76\\
    &NPSD-LSA \cite{li2016non} &1.28 &1.55 &1.87 &2.24 &2.58 &1.90 &0.64 &0.72 &0.79 &0.86 &0.91 &0.78\\
     &DNN-based \cite{tammen2020dnn}-LSA &1.62 &1.90 &2.20 &2.48 &2.73 & 2.19 &0.67 &0.75 &0.81 &0.85 &0.89 &0.79\\
    &Hybrid \cite{tao2023single}-LSA &1.36 &1.64 &1.91 &2.25 &2.58 &1.95 &0.65 &0.72 &0.77 &0.82 &0.86 &0.76\\
    &DNN \cite{tammen2020dnn} + Proposed System-LSA&\textbf{1.97} &2.10 &2.45 &2.76 &3.05 &2.47 &\textbf{0.73} &0.79 &0.85 &\textbf{0.90} &\textbf{0.94} &\textbf{0.84}\\
    &Proposed (BLSTM)-LSA &1.91 &\textbf{2.22} &\textbf{2.55} &\textbf{2.86} &\textbf{3.15} &\textbf{2.54} &0.72 &\textbf{0.80} &\textbf{0.86} &\textbf{0.90} &\textbf{0.94} &\textbf{0.84}\\
    &NSNet \cite{xia2020weighted} &1.66 &1.95 &2.22 &2.48 &2.72 &2.21 &0.69 &0.76 &0.83 &0.88 &0.91 &0.81\\
    &Proposed (Attention)-LSA &1.66 &1.95 &2.28 &2.60 &2.90 &2.28 &0.69 &0.76 &0.82 &0.88 &0.92 &0.81\\
    \midrule
    \multirow{7}*{\rotatebox{90}{NoiseX-92}}
    &Noisy &1.25 &1.52 &1.84 &2.18 &2.53 &1.86 &0.62 &0.72 &0.82 &0.89 &0.94 &0.80\\
    &UM-LSA \cite{gerkmann2011unbiased} &1.36 &1.70 &2.09 &2.48 &2.84 &2.09 &0.59 &0.70 &0.79 &0.87 &0.91 &0.77\\
    &NPSD-LSA \cite{li2016non} &1.44 &1.79 &2.16 &2.50 &2.81 &2.14 &0.61 &0.72 &0.81 &0.88 &0.93 &0.79\\
    &DNN-based \cite{tammen2020dnn}-LSA  & 1.41 & 1.77 & 2.14 & 2.48 & 2.75 &2.11 & 0.64 & 0.74 & 0.81 & 0.86 & 0.90 &0.79\\
    &Hybrid \cite{tao2023single}-LSA &1.31 &1.61 &1.95 &2.30 &2.63 &1.96 &0.62 &0.72 &0.79 &0.84 &0.87 &0.77\\
    &DNN \cite{tammen2020dnn} + Proposed System-LSA &1.57 &2.02 &2.45 &2.83 &3.17 &2.41 &\textbf{0.67} &0.77 &0.85 &0.91 &\textbf{0.95} &0.83\\
    &Proposed (BLSTM)-LSA &1.59 &\textbf{2.09} & \textbf{2.53} & \textbf{2.90} & \textbf{3.23} &\textbf{2.45} &\textbf{0.67} &\textbf{0.78} &\textbf{0.86} &\textbf{0.92} &\textbf{0.95} &\textbf{0.84}\\
    &NSNet \cite{xia2020weighted} &\textbf{1.73} &2.06 &2.35 &2.60 &2.83 &2.31 &0.64 &0.75 &0.83 &0.89 &0.92 &0.81\\
    &Proposed (Attention)-LSA  &1.47 &1.91 &2.39 &2.79 &3.13 &2.34 &0.65 &0.75 &0.84 &0.90 &0.94 &\textbf{0.82}\\
  \bottomrule
\end{tabular}
\label{Score}
\end{table*}

\begin{figure}[ht]
\centering
	\subfloat[Proposed-LSA ($\alpha_{snr}=0.98$)]{\includegraphics[width = 0.22\textwidth]{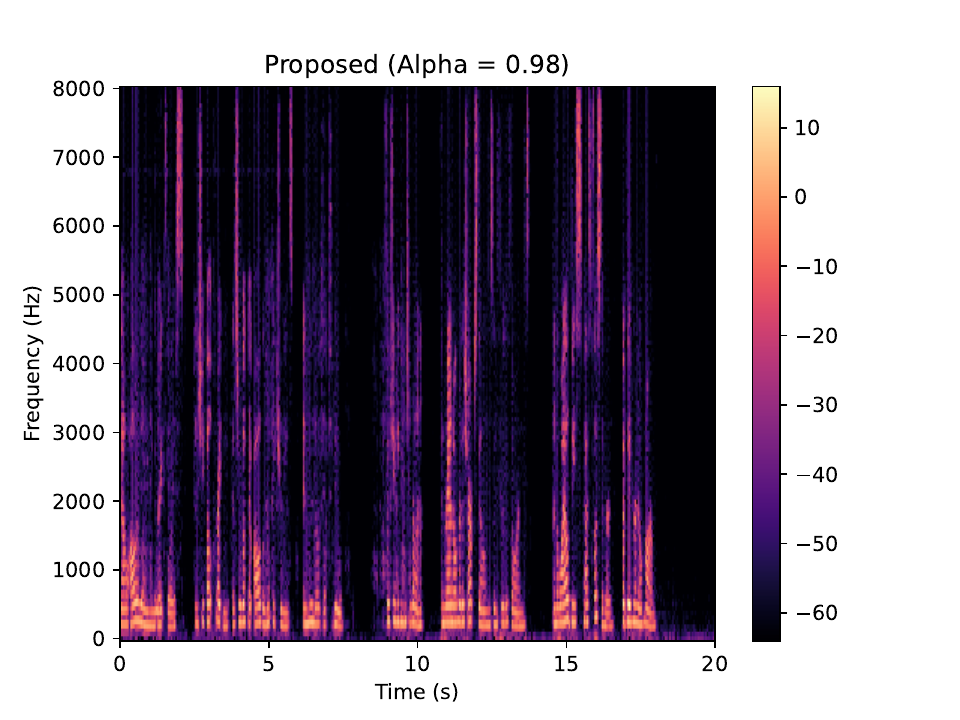}}
	\hfill
	\subfloat[Proposed-LSA ($\alpha_{snr}=0.90$)]{\includegraphics[width = 0.22\textwidth]{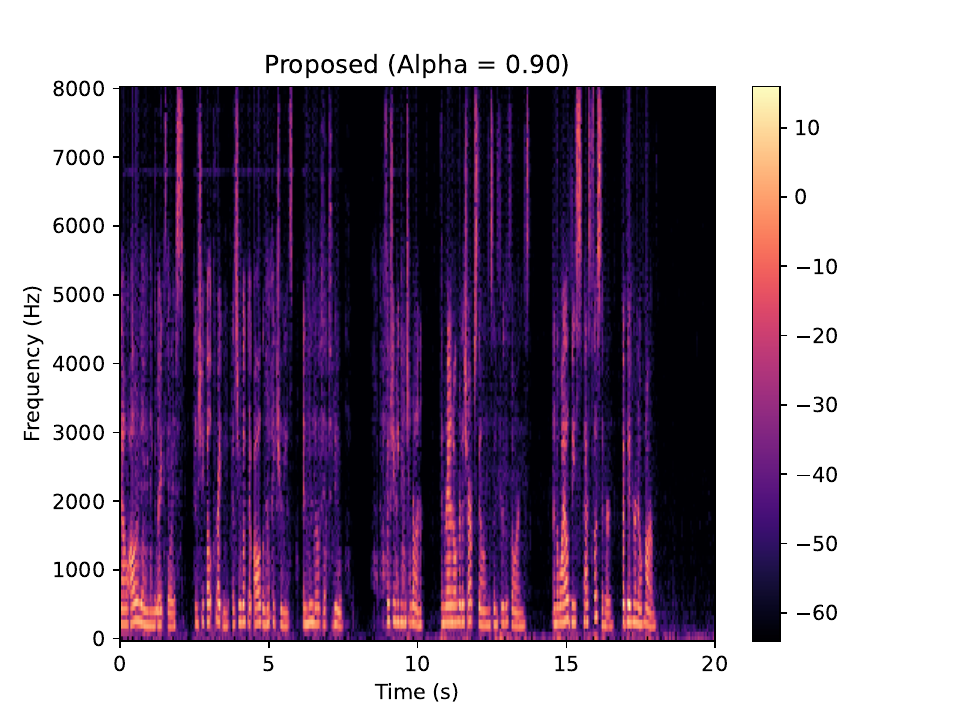}}
        \hfill
\caption{Two spectrograms of the enhanced speech by using our proposed (BLSTM) SPP estimator and sub-optimal MMSE-based noise PSD estimator with the LSA estimator, which have tuning factors of 0.98 and 0.90.}
\label{SE_results}
\end{figure}

Firstly, the one hour of testing utterance used in Section \ref{SPP performance} is also used for single-channel speech enhancement. Based on noise PSD estimation shown in Section \ref{noise_est}, two baselines of SPP estimators, UM and the DNN, are also integrated with LSA to enhance speech. For the baseline of the DNN-based SPP estimator, we also use our proposed sub-optimal MMSE to estimate noise PSD. To evaluate the performance of speech enhancement, SIG and BAK scores are computed by the objective evaluator DNSMOS \cite{reddy2021dnsmos} under different tuning factors $\alpha_{snr}$. Fig. \ref{dnsmos} (a) and (b) show SIG scores and BAK scores for different SPP-based speech enhancement frameworks. From the testing results, we can confirm that using the tuning factor can balance the speech distortion and noise reduction. Then, using our proposed method and based on the SPP estimate shown in Fig. \ref{SPP_EST} (d), the spectrogram of the enhanced speech with $\alpha_{snr}=0.90$ and $\alpha_{snr}=0.98$ are shown in Fig. \ref{SE_results}. Comparing Fig. \ref{SE_results} (a) and \ref{SE_results} (b), we can find that \ref{SE_results} (a) has more speech distortion than \ref{SE_results} (b). Moreover, when we change the tuning factor $\alpha_{snr}$ from 0.98 to 0.90, more speech components can be restored while there is also more residual noise.

From the above experimental results, we can find that the tuning factor $\alpha_{snr}$ in (\ref{eq23}) controls the trade-off between speech distortion and noise reduction. Additionally, \cite{malah1999tracking} and \cite{gerkmann2008improved} indicated that higher values of $\alpha_{snr}$ come with higher speech distortion and musical noise. However, higher residual noise exists when $\alpha_{snr}=0.90$, and more speech components can be restored to keep high speech quality. Therefore, to achieve high speech quality, in the following experiments, we set the tuning factor $\alpha_{snr}=0.90$ in (\ref{eq23}) to test the performance of speech enhancement, respectively, in terms of objective evaluation, i.e., PESQ and STOI.

In Table \ref{Score}, the PESQ and STOI scores are reported for various SPP-based speech enhancement frameworks and one end-to-end speech enhancement framework NSNet \cite{xia2020weighted}. For all SPP-based speech enhancement frameworks, the learning-based SPP estimation-directed speech enhancement achieved much higher speech enhancement performance than the statistics-based methods, i.e., UM \cite{gerkmann2011unbiased} and NPSD \cite{li2016non}. It can demonstrate that the learning-based method can estimate SPP from the background noise more accurately than the other reference statistics-based methods because the SPP is the key component of the speech enhancement framework. 

Additionally, for the learning-based SPP estimation-directed speech enhancement framework, the learning target for SPP estimation and noise PSD estimation approach also greatly impact speech enhancement performance. It can be demonstrated by the experimental results when we replace the learning target and noise PSD estimation approach by ours for the learning-based SPP estimation approach in \cite{tammen2020dnn}. Consequently, the PESQ and STOI scores were improved when using our proposed learning target and noise PSD estimation approach with the DNN proposed in \cite{tammen2020dnn}.

Finally, compared to the end-to-end approach NSNet, our proposed method is still effective even though at NoiseX92 testing dataset -10 dB, the PESQ score of our proposed method is lower than the NSNet, but the average score under all tested SNR levels is higher than NSNet. More importantly, our proposed method is more controllable and explainable than the end-to-end method, which means we can improve speech enhancement performance by achieving more accurate statistics estimation, i.e., SPP estimation accuracy in this paper and with the statistics estimate, the speech enhancement procedure can be controlled enabling provide one optimal solution for the trade-off between speech distortion and residual noise.

\subsection{Model Complexity Comparison}\label{complexity}
For computational complexity comparison, the number of parameters and FLOPs are computed for various speech enhancement systems.  As the learning-based SPP estimation-directed speech enhancement system, i.e., ours and the DNN-based in \cite{tammen2020dnn}, most of the model complexity comes from the SPP estimation part. We compute the number of parameters and FLOPs of the learning-based SPP estimation-directed speech enhancement systems based on the SPP estimation part. Additionally, since the statistics-based method has fewer parameters, the number of parameters of UM-based and NPSD-based speech enhancement systems were computed roughly. Table \ref{Complexity} shows the comparison of model complexity for various speech enhancement systems.
\begin{table}[ht]
  \centering
  \caption{Parameters and FLOPs comparison of the single-channel speech enhancement framework.}
  \scriptsize
  \begin{tabular}{ccc}
  \toprule
   Methods & Param & FLOPs (Mac/s, million) \\
  \midrule
  UM \cite{gerkmann2011unbiased}-LSA & $\le$ 600 & -\\
  NPSD \cite{li2016non}-LSA & $\le$ 600 & -\\
  Hybrid \cite{tao2023single}-LSA &0.04 M & 6.47\\
  DNN-based \cite{tammen2020dnn}-LSA & 1.39 M & 177.37\\
  NSNet \cite{xia2020weighted} & 1.26 M & 156.62 \\
  Proposed (BLSTM)-LSA &0.51 M & 65.68 \\
  Proposed (Attention)-LSA &0.23 M & 46.17 \\
  \bottomrule
\end{tabular}
\label{Complexity}
\end{table}

Compared to the learning-based speech enhancement system, statistics-based speech enhancement systems UM-LSA estimator and NPSD-LSA estimator have much fewer parameters and FLOPs, which means the statistics-based speech enhancement system has the lowest latency for real-time applications. The DNN-based LSA estimator has the highest parameters and FLOPs among the learning-based speech enhancement systems. Although the DNN-based LSA estimator also employs a DNN to estimate SPP, its model complexity is much higher than our proposed because our proposed learning-based SPP estimator is much lighter than the DNN-based SPP estimator in \cite{tammen2020dnn}. Importantly, based on Table \ref{Score} and Table \ref{Complexity}, we can confirm that our proposed SPP estimator not only achieved high SPP estimation accuracy but also had a low model complexity. From an application point of view, although the Hybrid-based method in \cite{tao2023single} achieved high SPP estimation accuracy and lower noise estimation error with the lowest model complexity among all compared learning-based SPP estimators, it cannot be deployed to speech enhancement. Moreover, our proposed method has much lower model complexity than the end-to-end approach NSNet. Therefore, in terms of speech enhancement performance and model complexity, our proposed speech enhancement system outperforms most existing SPP-based speech enhancement approaches.

\subsection{Performance of Multi-Channel Speech Enhancement}
In this subsection, we extend our proposed learning-based SPP estimator to multi-channel scenarios. Taking a simple case for multi-channel speech enhancement into account, we consider a linear microphone array with six microphones with a spacing of 0.1 m placed in a room. The room size is [10m, 8m, 3m] and the coordinate of the microphone array center is [5, 1.75, 1.7]. One sound source is considered, placed 1 to 3 meters away from the microphone center. To simulate realistic noise scenarios, we generate the diffuse noise by using the diffuse noise generator \cite{habets2007generating, habets2010comments}. The clean speech is taken from the DNS testing dataset and the noise is taken from the DNS and NoixeX-92 datasets. Finally, the multi-channel speech is generated with an input SNR ranging between -10 dB to 10 dB and a reverberation time of 0.2 seconds. For each noise dataset, 90 sets of data are simulated, and the simulation time is 20 seconds for each set.

The beamformer in \cite{anguera2007acoustic} and the additional post-filter are employed to process the multi-channel speech signal. Assuming that the beamformer already provides the optimum solution for the given sound field, following an additional post-filtering can significantly improve the SNR. Therefore, the single-channel solution can be extended to the multi-channel case. For post-filtering, the SPP estimator is integrated for multi-channel noise tracking and reduction in this paper. The LSA estimator, UM-LSA, our proposed method with LSA, and NSNet are employed as the post-filtering for the beamformer. To evaluate the multi-channel speech enhancement performance, PESQ, STOI, CSIG, and CBAK  scores are computed with the enhanced speech and the reference signal, which is the first microphone noise-free speech.
\begin{table}[ht]
  \centering
  \caption{Multi-channel speech enhancement performance comparison.}
  \scriptsize
  \begin{tabular}{cccccc}
  \toprule
  \multirow{1}*{} &\multirow{1}*{Methods} & \multicolumn{1}{c}{PESQ} & \multicolumn{1}{c}{STOI} & \multicolumn{1}{c}{CSIG} & \multicolumn{1}{c}{CBAK}\\
  \midrule
  \multirow{6}*{\rotatebox{90}{DNS}}
  &Noisy &1.59 & 0.72& 1.67& 1.55 \\
  &Beamformer \cite{anguera2007acoustic} &1.77 &0.75 &1.83 &1.38 \\
  &LSA \cite{ephraim1984speech} &1.70 &0.74 &1.54 &1.25 \\
  &UM \cite{gerkmann2011unbiased}-LSA &1.85 &0.74 &1.71 &1.39 \\
  &NSNet \cite{xia2020weighted} &2.15 &0.78 &1.79 &1.65 \\
  &Hybrid \cite{tao2023single}-LSA &1.83 &0.72 & 1.94 & 1.56\\
  &Proposed (BLSTM)-LSA &\textbf{2.34} &\textbf{0.80} &\textbf{2.53} &\textbf{1.81}  \\
  &Proposed (Attention)-LSA &2.12 &0.77 &2.21 &1.64  \\
    \midrule
  \multirow{6}*{\rotatebox{90}{NoiseX-92}}
  &Noisy &1.51 &0.68 &1.65 &1.44 \\
  &Beamformer \cite{anguera2007acoustic} &1.78 &0.73 &1.98 &1.31 \\
  &LSA \cite{ephraim1984speech} &1.83 &0.74 &1.83 &1.27 \\
  &UM \cite{gerkmann2011unbiased}-LSA &2.00 &0.72 &1.97 &1.39 \\
  &NSNet \cite{xia2020weighted} &2.22  &0.76 &2.06 &1.61 \\
  &Hybrid \cite{tao2023single}-LSA & 1.87 & 0.71 & 2.12 & 1.53 \\
  &Proposed (BLSTM)-LSA &\textbf{2.33} &\textbf{0.77} &\textbf{2.53} &\textbf{1.71}  \\
  &Proposed (Attention)-LSA &2.18 &0.76 &2.45 &1.66  \\
  \bottomrule
\end{tabular}
\label{Multi-Channel}
\end{table}

In Table \ref{Multi-Channel}, the comparison of the post-filtering performance for multi-channel speech enhancement is shown. As the additional post-filtering of the beamformer, most post-filters can further improve the speech quality, especially for our proposed method. Additionally, comparing all SPP-based post-filtering, i.e., LSA, UM-LSA, and our proposed method, our proposed method has great improvement in noise reduction and speech restoration which can demonstrate that our proposed method can also achieve high SPP estimation accuracy and noise PSD estimation accuracy in multi-channel scenarios. More importantly, in terms of the delay and the performance of post-filtering, our proposed method provides a more attractive solution for multi-channel speech enhancement.

\section{Conclusion}
This paper presents an appropriate learning target and noise PSD estimation approach for the learning-based $a$ $posteriori$ SPP estimation and SPP-directed noise PSD estimation. We argue that fixed $priori$ parameters of SPP may degrade the performance of learning-based SPP estimation. We therefore propose to use the adaptive parameters, i.e., the gain function of the optimal MMSE estimator and the actual $priori$ SNR to replace the fixed $a$ $priori$ SPP and SNR during the training procedure of SPP estimation. Additionally, since we use the adaptive parameters for the learning target, the noise PSD was estimated with the sub-optimal MMSE rather than the optimal MMSE, which can reduce the influence of the previous frame noise estimation error. From the noise PSD estimation evaluation results, we demonstrate that our proposed approach can further improve the performance of the learning-based SPP-directed noise PSD estimation. For speech enhancement, although we use a standard speech enhancement framework, the LSA estimator, we can also improve speech enhancement in terms of PESQ, STOI, CSIG, and CBAK. More importantly, in terms of both speech enhancement performance and model complexity, our proposed method outperforms most existing SPP-based speech enhancement approaches.

Firstly, the definition of the $a$ $posteriori$ SPP was given, and the learning target for the learning-based SPP estimator was presented. To improve the deep neural network capacity for acquiring realistic data representations, the parameters within the $a$ $posteriori$ were adjusted through observations derived from the training data, as opposed to relying on unchanging parameters, i.e., the fixed $a$ $priori$ SPP and SNR. As a result, the gain function of the optimal MMSE estimator as the $a$ $priori$ SPP and the actual $a$ $priori$ SNR were introduced. Subsequently, a light DNN model for $a$ $posteriori$ SPP estimation was proposed. The hybrid global-local information extraction component was incorporated into the DNN model, which contributes to reducing complexity and improving performance. Finally, to test the performance of the $a$ $posteriori$ SPP estimation, the distribution of the $a$ $posteriori$ SPP in the STFT domain was shown and two downstream tasks, i.e., noise PSD estimation and speech enhancement were performed. For the learning-based SPP estimator, the sub-optimal MMSE-based noise PSD estimation was proposed to reduce the error of the previous frame estimate, i.e., the noise PSD is computed by the $a$ $posteriori$ speech absence probability. For the extraction of clean speech from background interference, we utilized the noise PSD estimate with a standard speech enhancement framework employing the LSA estimator to reduce errors in noise PSD estimation.

\begin{backmatter}

\section*{Availability of data and materials}
Not applicable
  
\section*{Competing interests}
The authors declare that they have no competing interests.

\section*{Author's contributions}
All authors participate in methodology discussion, experimental design, and paper writing.


\bibliographystyle{bmc-mathphys} 
\bibliography{bmc_article,myabrv_new,my_reference,IEEEabrv}





\end{backmatter}
\end{document}